\documentclass{mem}
\usepackage{graphicx}
\begin{document}

\title{
The role of massive stars in the turbulent infancy of Galactic globular clusters: Feedback on the intracluster medium, and detailed timeline
}

%   \subtitle{}

\author{
 C. \,Charbonnel\inst{1,2}  
\and M. \,Krause\inst{1} 
\and T. \, Decressin\inst{1}
\and N. \,Prantzos\inst{3}
\and G. \, Meynet\inst{1}
          }
 
\institute{
Geneva observatory --
51, chemin des Maillettes,
1290 Versoix, Switzerland.
\email{Corinne.Charbonnel@unige.ch}
\and
IRAP, CNRS UMR 5277, Universit\'e de Toulouse, 14 avenue Edouard
               Belin, 31400 Toulouse Cedex 04, France
 \and
 Institut d'Astrophysique de Paris, CNRS UMR 7095, Universit\'e P.\&M.Curie, 98bis Bd Arago, 75014 Paris, France
}

\authorrunning{C.Charbonnel et al.}

\titlerunning{Massive stars in the infancy of globular clusters}

\abstract{ 
A major paradigm shift has recently revolutionized our picture of globular clusters (GC) that were long thought to be simple systems of coeval stars born out of homogeneous material.
Indeed, detailed abundance studies of GC long-lived low-mass stars performed with 8-10m class telescopes, together with high-precision photometry of Galactic GCs obtained with HST, 
have brought compelling clues on the presence of multiple stellar populations in individual GCs. 
These stellar subgroups can be recognized thanks to their different chemical properties (more precisely by abundance differences in light elements from carbon to aluminium; see Bragaglia, this volume) and by the appearance of multimodal sequences in the colour-magnitude diagrams (see Piotto, this volume). 
This has a severe impact on our understanding of the early evolution of GCs, and in particular of the possible role that massive stars played in shaping the intra-cluster medium (ICM) and in inducing secondary star formation. 
Here we summarize the detailed
timeline we have recently proposed 
for the first 40~Myrs in the lifetime of a typical GC following the general ideas of our so-called "Fast Rotating Massive stars scenario" (FRMS, Decressin et al. 2007b) and taking into account the dynamics of interstellar bubbles produced by stellar winds and supernovae.
More details can be found in Krause et al. (2012, 2013). \keywords{Globular clusters --
Stars: evolution -- ISM }
}
\maketitle{}

\section{Introduction}

Two main scenarios have been proposed in the literature to account for multiple stellar populations in GCs. 
They originally differ on the nature of the first generation (1G) stars potentially responsible for the GC self-enrichment as depicted from star-to-star abundance variations in light elements: polluters could indeed be either fast rotating massive stars (FRMS, e.g. Prantzos \& Charbonnel 2006; Decressin et al. 2007a,b;  Schaerer \& Charbonnel 2011; Krause et al. 2013) or AGB stars (e.g. Ventura et al. 2001, 2011), with very different consequences on the understanding of the dynamical evolution of GCs during the first 40-100 Myr of their life. 
Here we summarize the results presented in Krause et al. (2012, 2013) where we revisit the FRMS scenario for typical Galactic GCs. 

We consider as ``typical" a GC where no [Fe/H] abundance variation is observed from star to star,  like e.g.  NGC~6752.
According to Decressin et al. (2010), the initial mass of this GC must have been of the order of 9 $\times$10$^6$~M$_{\odot}$, with a half-mass radius of $\sim$3 pc, and a star formation efficiency of 30~\%.
Such a cluster must therefore have hosted $\sim$ 5700 1G massive stars (with masses above 25~M$_{\odot}$) under the assumption of a Salpeter IMF for 1G stars more massive than 0.8~M$_{\odot}$ (note that we assume a log-normal IMF for 1G and 2G stars of lower mass). 
We take into account the dynamics of interstellar bubbles produced by stellar winds and supernovae as well as the interactions of FRMS with the ICM and we describe the kinematics by a thin-shell model. 
We can then put forward a description of the time sequence for the first 40~Myrs of the evolution of GCs (the initial time being set at the assumed coeval birth of 1G stars) with three distinct phases. 

\section{Timeline for the first 40~Myrs in the lifetime of a typical GC}

\subsection{Wind bubble phase and 2G star formation}
\label{Windbubblephase}

During the first few Myrs after the formation of the 1G, the cluster 
is strongly impacted by the fast radiative winds of the numerous massive stars. 
Because their integrated energy is much smaller than the ICM's binding energy,
%Although 
those winds can not lift any noteworthy amount of gas out of the GC (see also Krause et al. 2012),  they will create large hot bubbles around massive stars and compress the ICM into thin filaments. 
Additionally in this environment ``classical" star formation is inhibited by the high Lyman-Werner flux (see also Conroy \& Spergel 2011) during the wind bubble phase as well as the subsequent supernova phase (\S~\ref{SNphase}).

However if massive stars have equatorial mass ejections that form decretion disks as we originally proposed in the FRMS scenario (Decressin et al. 2007b; see also Krti{\v c}ka et al. 2011) and as observed e.g. around Be-type stars (see e.g. Rivinius et al. 2001; Towsend et al. 2004; Haubois et al. 2012), accretion of pristine gas may proceed in the shadow of the equatorial stellar ejecta. 
Even if a substantial fraction of the power of the dominantly polewards-directed (because of rotation) fast wind would be channeled into the decretion disks, this would still be insufficient to unbind them from their parent stars. Similarly, close encounters with other stars at sufficiently small impact parameter are too unlikely to affect the disks significantly on an interesting timescale.
Second stellar generation (2G)  low-mass stars are then expected to form due to gravitational instability in the discs around 1G massive stars, those disks being fed both by the H-burning ashes ejected by the FRMS and by pristine gas. 

The total amount of gas that is made available to form 2G stars through this non-classical mode and with abundance properties similar to those observed today is high. 
Indeed Decressin et al. (2007b) models predict that FRMS lose about half of their initial mass via equatorial ejections loaded with H-burning products during the main sequence and the luminous variable phase.
As a consequence about $\sim$ 1~Myr after the birth of the 1G, the total mass lost through this mechanism by all the FRMS is of the order of 10$^4$~M$_{\odot}$/Myr. 
On the other hand a similar amount of pristine gas is expected to be brought by the accretion flow, with mixing proportions depending on the FRMS orbit-averaged accretion rate. 
2G stars will therefore form with the current light element abundances of the self-gravitating
disk of their massive 1G parent star, in agreement with the detailed abundance calculations of Decressin et al. (2007b) that reproduce nicely the observed  [O/Na] distribution in the typical GC NGC~6752.

The formation of 2G low-mass stars is expected to be complete after $\sim$ 9--10~Myrs (i.e., roughly the lifetime of a 25~M$_{\odot}$ star) after the formation of the 1G.
Note however that the duration of the wind bubble phase is rather uncertain since it depends on the mass limit (and therefore on the lifetime) at which stars explode as SNe or turn 
silently into black holes, which value is not settled yet (see Decressin et al. 2010 for references and further discussion on that issue).

\subsection{Supernova phase}
\label{SNphase}
Krause et al. (2012) showed that gas expulsion via supernovae explosions, which is generally invoked to drastically change the GC potential well and to induce substantial loss of 1G stars (Decressin et al. 2007a, 2010; Bekki et al. 2007; D'Ercole et al. 2008; Schaerer \& Charbonnel 2011; see also \S~\ref{darkremnantphase}), does not work for typical GCs. 
Indeed, while the energy released by SNe usually exceeds the binding energy, SNe-driven shells turn out to be destroyed by 
the
Rayleigh-Taylor instability before they reach escape speed; therefore the shell fragments that contain the gas remain bound to the cluster. 

Although the SNe-ejecta will likely be mixed with the cold phase of the ICM, they are not expected to appear in the composition of 2G stars. 
Indeed once the first very energetic SNe explode, the ICM becomes highly turbulent 
and, since the Bondi-accretion rate depends on $v^3$, where $v$ is the relative velocity between the star and the local ICM, 
can not be accreted anymore onto FRMS decretion disks nor (yet) onto any dark remnants  that are being continuously produced (i.e., stellar black holes from stars initially more massive than 25~M$_{\odot}$ stars and 
later
neutron stars from initially less massive stars). 
The SNe phase is therefore expected to be sterile as far as star formation is concerned.

\subsection{Dark remnant accretion phase}
\label{darkremnantphase}

The FRMS and the AGB scenario strongly differ on many aspects of the early dynamical and chemical evolution of the young GCs. 
They both agree however on the fact that GCs must have been much more massive initially than today and call for substantial  loss of 1G low-mass stars  in order to explain todays observed proportions between 1 and 2G stars (Carretta et al. 2010; Schaerer \& Charbonnel 2011). 
Within the framework of the AGB cooling flow scenario, this powerful event is expected to occur before the formation of 2G stars.
At the opposite, 2G stars are expected to form much earlier in the FRMS scenario as described in \S~\ref{Windbubblephase}. 

Krause et al. (2012) proposed that very fast gas expulsion, which is expected to unbind a large fraction of 1G low-mass stars sitting initially in the GC outskirts, may be powered by the sudden and concomitant activation of 1G dark-remnants (i.e., 1.5~M$_{\odot}$ neutron stars from stars with masses between 10 and 25~M$_{\odot}$, and stellar mass black holes  
with 3~M$_{\odot}$ from more massive stars).  
Accretion of interstellar gas onto these objects can set in only once turbulence has sufficiently decayed in the ICM after the type II SNe have ceased, i.e. about 35-40~Myrs 
(corresponding to the lifetime of a 9~M$_{\odot}$ star) after the coeval birth of 1G stars. 
Assuming coherent onset of accretion onto all the 1G stellar remnants and considering that they could contribute to gas energy at 20$\%$ of their Eddington luminosity, Krause et al. (2012) showed that a sufficient amount of energy is released in at most 0.06~Myrs (i.e., about half a crossing time), which is short enough to avoid gas fall back due to Rayleigh-Taylor instability and to eject the ICM together with the outer weakly bound 1G low-mass stars.

After the major expulsion event described above, the cluster is expected to be devoid of pristine gas.
However the dark remnants may be reactivated later, this time through the eventual accretion of slow-AGB winds. 
When this happens we expect again quick gas expulsion, which would inhibit any subsequent star formation from the AGB ejecta. 
This situation is alike the possible activation of super-massive black holes in nuclear star clusters and elliptical galaxies by AGB-winds (e.g. Gaibler et al. 2005; Davis et al. 2007; Schartmann et al. 2010).

Importantly, Krause et al. (2012) find a limiting initial cloud mass of $\sim$ 10$^7$~M$_{\odot}$ for the dark remnant scenario to work. 
Above that mass, the cold pristine gas may not be ejected, and subsequent star formation may proceed in the ICM that would then be enriched in SNe ejecta. 
This can explain the fact that the more massive (and atypical) GCs (Omega Cen, M22, M54, NGC 1851, NGC 3201; Carretta et al. 2010, 2011; Johnson \& Pilachowski 2010; Marino et al. 2011; Simmerer et al. 2012, 2013) do exhibit internal [Fe/H] spread, while less massive (typical) ones do not. 

\section{Conclusions and perspectives}
After the tremendous paradigm shift that has recently revolutionized our view of GCs, 
new observational and theoretical advances make the present a particularly exciting time for the study of the origin, formation, and evolution of these objects in the Milky Way as well as in external galaxies. 

On the theoretical front one of the fundamental issues concerns the physics of the interstellar medium and its interactions with stars.
Also, the formation of 2G stars in massive disks fed by both equatorial ejections of parent FRMS and accretion of pristine gas has to be studied in details.
Cutting-edge strategies have to be established in order to tackle the open issues from an innovative point of view. 
In particular, relevant multi-dimensional MHD simulations would greatly help in clarifying details of chemical and dynamical feedback of 1G massive stars and of secondary star formation. 
Work is in progress in that direction.

\begin{acknowledgements}
We acknowledge support from the Swiss National Science Foundation (FNS), 
from the cluster of excellence ``Origin and Structure of the Universe'', and the ESF
EUROCORES Programme ``Origin of the Elements and Nuclear History of the Universe".
\end{acknowledgements}

\bibliographystyle{aa}

\end{document}